\documentclass[12pt]{iopart}

\usepackage{iopams}  
\usepackage{hyperref}
\usepackage{dcolumn,bm}
\usepackage{graphicx}
\usepackage{multirow}
\usepackage{epstopdf}
\usepackage[table]{xcolor}
\usepackage{array}
\usepackage{longtable}

\newcommand{\be}{\begin{equation}}
\newcommand{\ee}{\end{equation}}
\newcommand{\ba}{\begin{eqnarray}}
\newcommand{\ea}{\end{eqnarray}}

\newcommand{\mev}{\, \rm{MeV}}

\newcommand{\hw}{\hbar \Omega}

\newcommand{\Nmax}{N_{\rm max}}
\newcommand{\Neff}{N_{\rm eff}}
\newcommand{\Leff}{L_{\rm eff}}

\newcommand{\eqref}[1]{(\ref{#1})}

\begin{document}

\title{Infrared extrapolations for atomic nuclei}

\author{R.~J.~Furnstahl$^{1}$, G.~Hagen$^{2,3}$, T.~Papenbrock$^{3,2}$, and K.~A.~Wendt$^{3,2}$}

\address{$^1$Department of Physics, The Ohio State University,
  Columbus, OH 43210}

\address{$^2$Physics Division, Oak Ridge National Laboratory, Oak
  Ridge, TN 37831 USA}

\address{$^3$Department of Physics and Astronomy, University of
  Tennessee, Knoxville, TN 37996, USA}

\eads{\mailto{furnstahl.1@osu.edu}, \mailto{hageng@ornl.gov}, 
    \mailto{tpapenbr@utk.edu}, \mailto{kwendt2@utk.edu}}

\begin{abstract}
  Harmonic oscillator model-space truncations introduce systematic
  errors to the calculation of binding energies and other observables.
  We identify the relevant infrared scaling variable and give values
  for this nucleus-dependent quantity.  We consider isotopes of oxygen
  computed with the coupled-cluster method from chiral nucleon-nucleon
  interactions at next-to-next-to-leading order and show that the
  infrared component of the error is sufficiently understood to permit
  controlled extrapolations. By employing oscillator spaces with
  relatively large frequencies, well above the energy minimum, the
  ultraviolet corrections can be suppressed while infrared
  extrapolations over tens of MeVs are accurate for ground-state
  energies.  However, robust uncertainty quantification for
  extrapolated quantities that fully accounts for systematic errors is
  not yet developed.
\end{abstract}


\submitto{\JPG}
\maketitle


\section{Introduction}

Wave-function based methods for computing atomic nuclei, such as exact
diagonalization~\cite{navratil2009,barrett2013}, coupled
cluster~\cite{kuemmel1978,bishop1991,hagen2013c}, self-consistent
Green's functions~\cite{dickhoff2004}, or in-medium similarity
renormalization group (SRG)~\cite{tsukiyama2011,hergert2013}, require
a computational effort that depends strongly on the size of the 
Hilbert space used.  Truncating this model space introduces systematic
errors that must be quantitatively analyzed for a given Hamiltonian
and nucleus.
By understanding these errors, we can devise controlled extrapolation
techniques for energies and other observables.

\begin{figure}[htb]
\begin{center}
\includegraphics[width=5in]{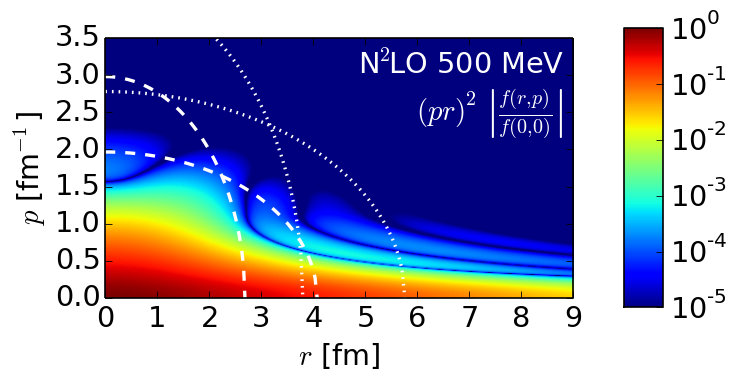}\\
\includegraphics[width=5in]{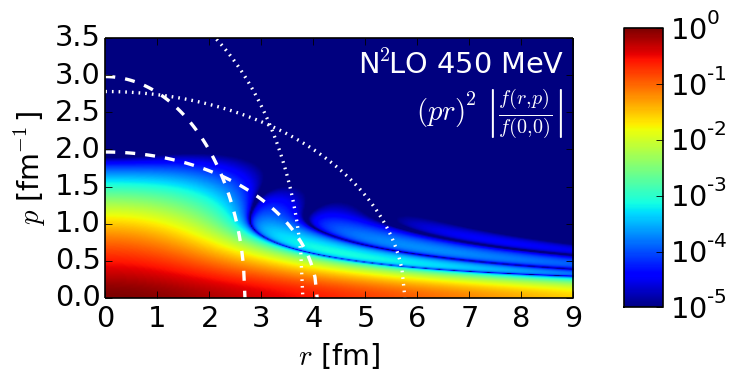}
\end{center}
\caption{(Color online) Top: \lq\lq{}$s$-wave\rq\rq{} projection of
  Wigner function from deuteron wave function for a chiral N$^2$LO
  potential with a regulator cutoff of $\Lambda_{\rm UV}=500\mev$.
  Abscissa is $r$ in units of fm, ordinate is momentum $p$ in units of
  fm$^{-1}$.  The dashed lines are semi-classical harmonic oscillator
  cutoffs for $N=4$ and two different oscillator frequencies ($\hbar\Omega=20$ and
  $50\mev$).  The dotted lines are the same with $N=8$. Bottom: same as
  top but for a lower regulator cutoff $\Lambda_{\rm UV}=450\mev$.}
  \label{fig1} 
\end{figure}

Though we deal with a quantum mechanical problem, it is most
instructive to look first at classical phase space.
Let us consider the deuteron wave function, calculated from a
realistic two-body interaction, and compute its Wigner transform.  We
recall that the Wigner transform
\be
  f(\bm{r},\bm{p}) = \int d^3\bm{k}\, 
  \frac{e^{\imath \bm{r}\cdot \bm{k}}}{(2\pi)^3}
  \psi^*(\bm{p}+\bm{k}/2)\psi(\bm{p}-\bm{k}/2) \;,
\ee 
is a mapping of a wave function to a phase-space distribution.
Figure~\ref{fig1} shows the results for chiral effective field theory
(EFT) Hamiltonians~\cite{ekstrom2014} with two different cutoffs.  We
see that the dominant part of the Wigner transform extends in (radial)
position essentially up to the deuteron radius (though the exponential
tail casts a long shadow), and in momentum space up to the cutoff of
the interaction.

Generalizing this result, the number of single-particle states
required to compute a nucleus with radius $R$ from an interaction with
cutoff $\Lambda$ is essentially given by
\be 
\label{nsp}
n={1\over (2\pi)^3} \int\limits^R d^3 r \int\limits^\Lambda d^3p
\propto (R\Lambda)^3 \;.  
\ee
Here, we have not counted spin/isospin degrees of freedom. The key
result is that the number of single-particle states grows as the third
power of the cutoff, and it is proportional to $R^3\propto A$ for a
nucleus with mass number $A$. The proportionality constant in
Eq.~(\ref{nsp}) depends on the actual basis one uses, and some
efficiencies can possibly be gained from abandoning the oscillator
basis in favor of Berggren bases~\cite{michel2009}, Sturmian
bases~\cite{caprio2012} or discrete variable
representations~\cite{bulgac2013}.  Equation~(\ref{nsp}) also makes
clear that halo nuclei can be as expensive to compute as much heavier
nuclei, and that much is to be gained by lowering the cutoff $\Lambda$
of the interaction. This latter point explains why low-momentum
interactions (e.g.\ $V_{{\rm low}\,k}$~\cite{bogner2003} or
SRG~\cite{bogner2007}) are so popular.

When working with the harmonic-oscillator basis, its extent in
position and momentum space must exceed the radius of the nucleus and
the cutoff of the interaction, respectively. In other words, one must
choose the oscillator frequency and the maximum excitation energy such
that the ellipsoidal phase-space area of the oscillator covers the
relevant parts of the Wigner function for the nucleus. Let
$b=\hbar/\sqrt{m\hbar\Omega}$ denote the oscillator length, where $m$
is the nucleon mass. Then a simple semiclassical argument implies that
for an oscillator basis with up to $N$ excited quanta, we must demand
\ba 
\label{semiclass}
\sqrt{2N} \ge R/b \quad\mbox{and}\quad \sqrt{2N} \ge \Lambda b \;.
\ea
The ellipses corresponding to equality signs in these expressions are
also shown in Fig.~\ref{fig1}.
We see that only the tails of the Wigner function -- both in the
directions of position and momentum -- extend beyond the ellipses.
Let us also note that Fig.~\ref{fig1} serves mainly as an illustration
but should not be used for quantitative conclusions. Only in the
semiclassical limit is it possible to quantitatively relate Wigner
functions to classical phase-space structures~\cite{bohigas1993}. The
questions to address are what error results from the omission of these
tails for the binding energy and other observables in a finite
oscillator basis, and how quickly do these quantities converge as the
number of basis states is increased?

\section{Infrared cutoffs of the oscillator basis}

The semi-classical estimates~(\ref{semiclass}) are a useful guide, but 
for practical purposes we
have to understand how the infrared (IR) cutoff $\Lambda_{\rm
  IR}\equiv \pi/L$ and ultraviolet (UV) cutoff $\Lambda_{\rm UV}$ of
a finite oscillator basis can become tools for the extrapolation of
observables. The leading-order estimates for these cutoffs are based
on $L\approx\sqrt{2N} b$ and $\Lambda_{\rm UV}\approx\sqrt{2N}/b$,
respectively~\cite{stetcu2007,hagen2010b,coon2012}.  Clearly, $L$ and
$\Lambda_{\rm UV}$ are semiclassical estimates for the extent of the
oscillator basis in position and momentum space, respectively.

The nucleon-nucleon forces (and many-body forces)
used for wave-function based methods, such as chiral effective field theory
(EFT) or low-momentum interactions, all have momentum-space regulators that
rapidly drive them to zero above a cutoff $\Lambda$.  
If $\Lambda_{\rm UV}$ is less than or comparable to $\Lambda$, the UV corrections
to the energy and other observables will depend on the details of the regulators
(but less on the nucleus under consideration).  But when $\Lambda_{\rm UV}$
sufficiently exceeds $\Lambda$ (see Sec.~\ref{sec:results}), 
rapid UV convergence is observed in practical calculations,
so that UV extrapolations are not needed.

In this paper, we focus solely on IR extrapolations. We note that the long
wavelength structure of a bound state stems from its exponential
fall-off in position space and is independent of the details of the
interaction. A finite extent of the oscillator basis in position
space cuts off this exponential tail and thereby yields an energy
correction. References~\cite{furnstahl2012,more2013,furnstahl2014}
provide a theoretical basis for IR extrapolations. 
It is established in these papers
that a finite oscillator basis with $N$ excited quanta
and oscillator length $b$ is, at low momenta, indistinguishable 
from a box of size
\be
\label{L2}
L_2(N,\hbar\Omega)\approx\sqrt{2(N+3/2+2)} b \;.
\ee
Here, the approximate sign indicates that this is the next-to-leading
order approximation of the box size in the limit of $N\gg 1$. This
result can be derived from the fact that $\Lambda_{\rm IR}^2\equiv
(\pi/L_2)^2$ is the lowest eigenvalue of the momentum operator squared
in the finite oscillator basis~\cite{more2013}. A single-particle wave
function of an $s$-wave bound state, approximated in a finite
oscillator basis, only differs by high-momentum components from the
same wave function approximated in a spherical well of radius $L_2$.
For a partial wave with angular momentum $l$, one has $N=2n+l$ in
Eq.~(\ref{L2}). Thus, for fixed oscillator spacing $\hbar\Omega$ the
length $L_2(N,\hbar\Omega)$ is a staircase function that increases at even
(odd) values of $N$ for even (odd) values of $l$.

This knowledge can be used to understand the IR extrapolations of
bound states. The equivalent finite size $L_2$ of the oscillator basis 
has the effect on a position-space bound state of
enforcing that the wave function has a node at $L_2$.
This correction to the true bound-state wave function is a 
long-wavelength phenomenon and can thus be understood in model-independent
ways~\cite{more2013,furnstahl2014}. The leading-order IR extrapolation
formula for an $s$-wave bound-state energy of a single-particle system
is
\be
\label{master}
E(L_2) = E_\infty +  {\hbar^2 k_\infty
  \gamma_\infty^2\over  \mu} \exp{\left(-2k_\infty L_2\right)}  \;.  
\ee
Here, $k_\infty$ is the bound-state momentum, i.e.
$E_\infty=\hbar^2 k_\infty^2/(2\mu)$ is the bound-state energy, $\mu$
is the reduced mass, and $\gamma_\infty$ is the asymptotic
normalization constant. Higher-order corrections, which are suppressed
by powers of $\exp{\left(-2k_\infty L_2\right)}$,
and extensions to
general orbital angular momentum are also known~\cite{furnstahl2014}.

Although the IR extrapolation formula~(\ref{master}) was derived for a
single-particle degree of freedom (or systems that can be reduced to
such), it has also been applied to bound states of many-body
problems~\cite{furnstahl2012,jurgenson2013,roth2013,soma2013,hergert2013b,saaf2014} via 
\ba
\label{master2}
E(L) &=& E_\infty + A_\infty \exp{\left(-2k_\infty L\right)} \;,
\ea
with $L(N,\hw)$ still to be specified.  The idea is that $k_\infty$
can be generally interpreted as the (least) separation energy for the
nucleus under consideration, so that (\ref{master2}) follows from the
two-body derivation in Ref.~\cite{furnstahl2014} but now using the $S$
matrix for the corresponding break-up reaction.  Because of the many
approximations involved, $A_\infty$, $E_\infty$, and $k_\infty$ are
all treated as fit parameters.  To apply the extrapolation
formula~(\ref{master2}), one needs to work with bound states (i.e. the
fully converged energy needs to be negative), $L$ must exceed the
radial extent of the nucleus under consideration, and the UV cutoff
$\Lambda_{\rm UV}$ must be sufficiently greater than the cutoff of the
interaction.  In practice, one can calculate using large values of
$\hbar\Omega$ to ensure that UV corrections are much smaller than the
IR corrections.

\begin{figure}[t]
\begin{center}
\includegraphics{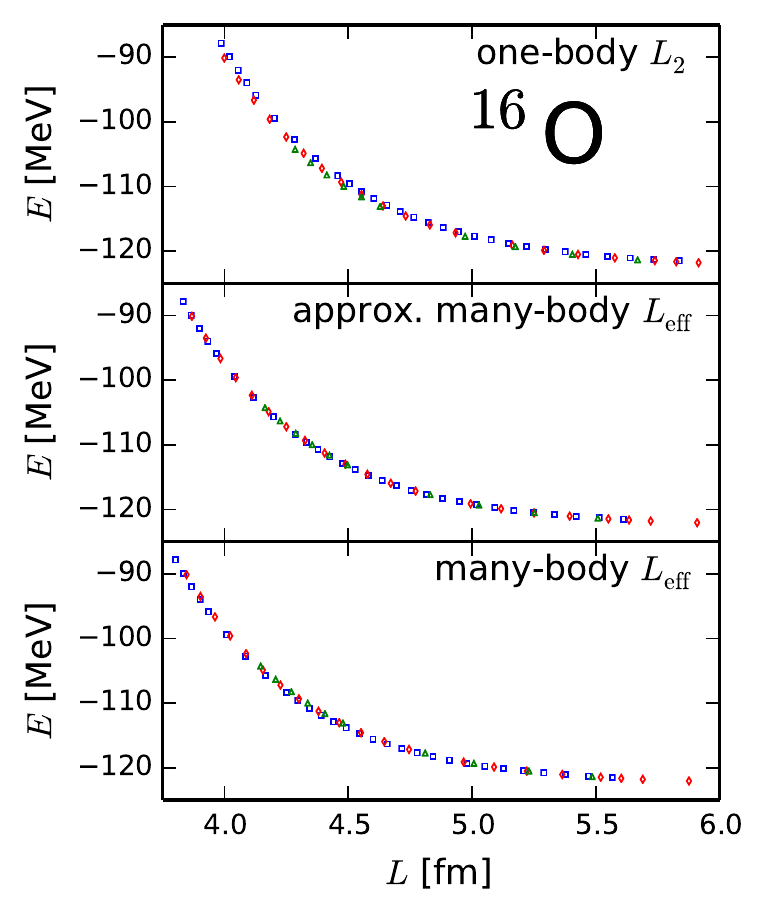}
\includegraphics{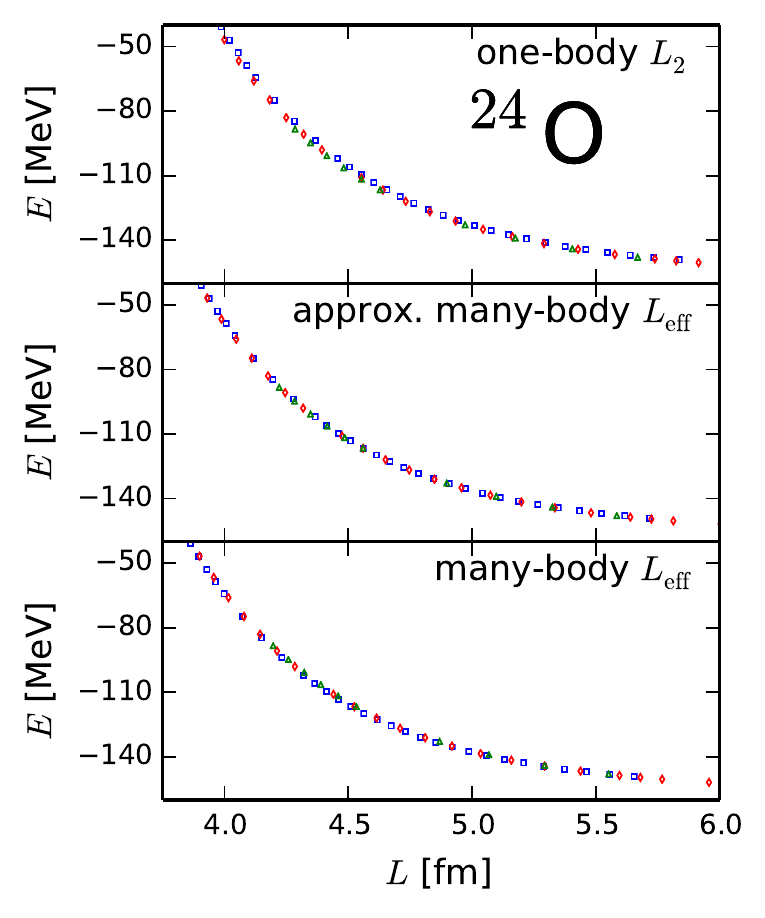}
\end{center}
\caption{Ground-state energies versus the scaling variable $L$ from
  coupled-cluster theory in the CCSD approximation for $^{16}$O (left)
  and $^{24}$O (right) using a chiral N$^2$LO potential with a
  regulator cutoff of $\Lambda_{\rm UV}=500\mev$.  The squares,
  diamonds and triangles stem from model spaces with $N=8,10,12$,
  respectively. The top panels use the naive single-particle choice $L
  =L_2(N,\hw)$, while the middle and bottom panels use approximate and
  exact $L = \Leff(N,\hw)$, respectively, as described in
  Sec.~\ref{sec:leff}.}
  \label{fig:scaling} 
\end{figure}

Let us consider applying Eq.~(\ref{master2}) to compute the
ground-state energies of oxygen isotopes. As a first step, we 
plot the energies obtained using different model spaces as a function of
the effective box size $L$. The top panels in Fig.~\ref{fig:scaling} show the
results for $^{16}$O (left) and $^{24}$O (right)
when plotted versus $L=L_2$. The data points stem from model spaces with
$N=8,10,12$ oscillator quanta and $\Lambda_{\rm UV} > 750\,$MeV, so that
UV corrections are presumably small. Clearly, the data points do not
fall on a single line for either nucleus, and similar results are found 
for $^{22}$O. Earlier studies~\cite{more2013} showed that the smoothness
of $E(L)$, when plotted for a large set of $N$ and $\hw$ values for
which UV corrections are small, is a sensitive diagnostic of the
quality of $L(N,\hw)$. Thus, $L=L_2$ is not the accurate box size of
the harmonic oscillator basis for fermionic many-body systems.  In the
next section we derive a more accurate value $\Leff$ by revisiting
the underlying basis for the effective box size. As we will see, this
value depends on the nucleus under consideration.

\section{Box size $\Leff(N,\hw)$ for nuclei}
\label{sec:leff}

The paradigm for IR extrapolations in the harmonic oscillator basis
can be stated as follows: {\it For sufficiently low energies (and long
  wavelengths), a finite oscillator basis is indistinguishable (in the
  sense of an effective theory) from a spherical box with an
  appropriately chosen radius.}  This radius is determined by matching
the lowest (most infrared) eigenvalue of the squared momentum operator
in the oscillator basis to the lowest value in the box.  To apply this
matching procedure to the many-body case, we have to consider the
lowest \emph{total} squared momentum $\sum_{i=1}^A p_i^2$ for a given
nucleus.  To do so, we identify the occupation numbers $\nu_{nl}$ that
give the lowest kinetic energy for that nucleus, and then equate the
eigenvalue of the total squared momentum operator for the oscillator
basis to the total squared momentum for the same number of nucleons
(occupying identical partial waves) in a spherical box of size
$\Leff$, i.e. 
\be
\sum_{nl} \nu_{nl} \kappa_{ln}^2 = \sum_{nl} \nu_{nl} 
  \left({a_{l,n}\over \Leff}\right)^2  \;. 
\ee
Here, $\kappa_{ln}^2$ is the eigenvalue of the single-particle squared
momentum operator and 
$a_{l,n}$ is the $(n+1)^{\rm th}$ zero of the spherical Bessel function 
$j_l$~\cite{Olver:2010:NHMF}.
This yields
\be
\label{leff}
\Leff = \left(\frac{\sum_{nl} \nu_{nl} a_{l,n}^2 }{\sum_{nl} \nu_{nl} \kappa_{ln}^2} \right)^{1/2} \;. 
\ee
We can parameterize this result using $L_2$
by introducing $\Neff$, which is defined as a function of $N$ by
\be
    \Leff(N,\hw) \equiv L_2(\Neff,\hw) \;.
\ee
We note that the approach to
the many-body scaling variable $\Leff$ can be extended to any localized basis
set by numerically computing the eigenvalues of the total squared
momentum operator and equating them to the corresponding eigenvalues
of a spherical box with radius $\Leff$.

For an understanding of $\Leff$ (or $\Neff$), it is useful to consider
analytical approximations to $\Leff$. These are based on approximate
expressions for the eigenvalues $\kappa_{lm}$ which are valid for
model spaces with $N\gg 1$.  In the single-particle problem (or the
deuteron in its center-of-mass frame), the $(m+1)^{\rm th}$ eigenvalue
$\kappa_{lm}^2$ in a harmonic-oscillator model space with up to $N$
quanta of excitation is given to good
approximation~\cite{furnstahl2014} by
\be
\label{kappal}
\kappa_{lm}^2 \approx \frac{a_{l,m}^2}{2(N_l +3/2 +2)b^2} = \frac{a_{l,m}^2}{\left(L_2(N,\hw)\right)^2} \;, 
\ee
and 
\ba
\label{Nl}
N_l =\left\{
\begin{array}{ll}
N \ , & \mbox{for $N$ and $l$ even or for $N$ and $l$ odd;}\\
N-1 \ , & \mbox{for $N$ even and $l$ odd or for $N$ odd and $l$ even.}\\
\end{array}\right.
\ea
As a check, consider a single particle in a model space with $N =
2n+l$.  We equate the lowest eigenvalue $\kappa_{l0}^2$ to the lowest
eigenvalue $(a_{l,0}/L)^2$ in a spherical box of radius $L$ for
angular momentum $l$. This yields $L = L_2(N,\hw)$ as defined in
Eq.~\eqref{L2}.

Thus, approximate values for $\Leff$ are found by inserting the
analytical approximations~(\ref{kappal}) into Eq.~(\ref{leff}). The
middle panels of Fig.~\ref{fig:scaling} show the ground-state energies
for $^{16,24}$O when plotted as a function of this approximate value
for $\Leff$. Compared to the top panel ($L=L_2$), the improvement is
considerable and clearly visible. The bottom panels of
Fig.~\ref{fig:scaling} show the ground-state energies for $^{16,24}$O
when plotted as a function of the exact numerical value for $\Leff$.
Again, the data points fall close to a single line. For reasons we do
not yet understand, the approximate $\Leff$ (middle panel) leads to
the smoothest line in $^{16}$O, while the exact $\Leff$ yields the
smoothest line in $^{24}$O.

Table~\ref{tabNeff} gives $\Neff$ for isotopes $^{12,14}$C and
$^{16,22,24}$O. These results are determined by diagonalizing the
operator $\sum_{i=1}^A p_i^2$ in the specified model space. We note
that the ``square'' model space (defined by each particle allowed up
to $N$ oscillator quanta) is used in the coupled-cluster calculations
of this paper and by other
methods~\cite{dickhoff2004,tsukiyama2011,hergert2013}. This model
space differs from the ``triangular'' full $N\hbar\Omega$ model space
used in most NCSM calculations.

\begin{table}[hbtp]
\setlength{\tabcolsep}{2ex}
\begin{center}
  \begin{tabular}{r|rrrrr}
    \hline\hline 
    & \multicolumn{5}{c}{$N_{\rm eff}$} \\ 
\multicolumn{1}{c|}{$N$} &
\multicolumn{1}{c}{$^{12}{\rm C}$} &
\multicolumn{1}{c}{$^{14}{\rm C}$} &
\multicolumn{1}{c}{$^{16}{\rm O}$} &
\multicolumn{1}{c}{$^{22}{\rm O}$} &
\multicolumn{1}{c}{$^{24}{\rm O}$} \\
\hline
 1 &  0.365  &  0.393  &  0.413  &  \multicolumn{1}{c}{---}  &  \multicolumn{1}{c}{---}  \\
 2 &  0.745  &  0.709  &  0.684  &  0.899  &  0.939  \\
 3 &  2.515  &  2.545  &  2.566  &  1.939  &  1.844  \\
 4 &  2.902  &  2.867  &  2.842  &  3.092  &  3.137  \\
 5 &  4.586  &  4.617  &  4.639  &  4.116  &  4.033  \\
 6 &  4.976  &  4.941  &  4.917  &  5.187  &  5.236  \\
 7 &  6.629  &  6.660  &  6.682  &  6.205  &  6.127  \\
 8 &  7.020  &  6.985  &  6.961  &  7.245  &  7.296  \\
 9 &  8.658  &  8.689  &  8.711  &  8.259  &  8.185  \\
10 &  9.049  &  9.015  &  8.990  &  9.285  &  9.337  \\
11 & 10.678  & 10.709  & 10.732  & 10.297  & 10.225  \\
12 & 11.070  & 11.036  & 11.011  & 11.313  & 11.366  \\
13 & 12.693  & 12.725  & 12.748  & 12.324  & 12.254  \\
14 & 13.085  & 13.051  & 13.027  & 13.335  & 13.389  \\
15 & 14.705  & 14.737  & 14.760  & 14.345  & 14.276  \\
16 & 15.097  & 15.064  & 15.040  & 15.352  & 15.406  \\
17 & 16.715  & 16.747  & 16.770  & 16.361  & 16.293  \\
18 & 17.107  & 17.074  & 17.050  & 17.366  & 17.421  \\
19 & 18.723  & 18.755  & 18.778  & 18.375  & 18.307  \\
20 & 19.115  & 19.082  & 19.058  & 19.377  & 19.433  \\
    \hline\hline
  \end{tabular}
\end{center}
\caption{Effective excitation number $N_{\rm eff}$ for isotopes $^{12,14}$C 
  and $^{16,22,24}$O, computed from the exact eigenvalues of the total 
  squared momentum operator. Where no result is given, the model space is too small to accommodate $A$ nucleons.}
\label{tabNeff}
\end{table}

Let us make some comments on the values of $\Neff$ in
Table~\ref{tabNeff}. We recall that $N_{\rm eff}$ is computed from the
exact eigenvalues $\kappa_{lm}^2$ of the total squared momentum
operator. The resulting values for $N_{\rm eff}$ are a few percent
smaller than the results one obtains from using the $N\gg 1$
approximation~(\ref{kappal}) in the computation of $N_{\rm max}$.
However, these latter approximations $N^{\rm approx}_{\rm eff}$ can be
understood semi-quantitatively as follows.  For the nucleus $^{4}$He
(not shown) we have $N^{\rm approx}_{\rm eff}=N$ ($N^{\rm approx}_{\rm
  eff}=N-1)$ for even (odd) $N$ because only $s$-waves are occupied,
so $L_2(N,\hbar\Omega)=\Leff(N,\hbar\Omega)$.  For $^{12,14}$C and
$^{16}$O, $p$-waves are at the Fermi surface, and for pure $p$-waves
and even (odd) $N$ one has $N^{\rm approx}_{\rm eff}=N-1$ ($N^{\rm
  approx}_{\rm eff}=N$), see Eq.~(\ref{kappal}). Thus, our values for
$N^{\rm approx}_{\rm eff}$ are closest to these extreme values for
$^{16}$O and somewhat farther away for $^{14,12}$C.  For $^{22,24}$O,
$d$-shell orbitals are at the Fermi surface for the kinetic energy,
countering the effects from the lower-lying $p$ waves and effectively
increasing (decreasing) $N^{\rm approx}_{\rm eff}$ for even (odd) $N$.
As naively expected, the effect is larger in $^{24}$O than in
$^{22}$O.  From these arguments, one expects $N-1\le N^{\rm
  approx}_{\rm eff} \le N$ in general.

We note that the defect $N-N_{\rm eff}$ varies very little (focusing
on either even or odd $N$) as $N$ varies. In practice, one might thus
take a fixed defect (say from $N=10$ or so) and consider the small
variation of the defect a higher-order correction. We also note an
odd-even staggering of $N_{\rm eff}$.  This staggering is caused by
Eq.~(\ref{Nl}) and has its root in the fact that $L_2$ increases for
even (odd) values of angular momentum only at even (odd) values of
$N$. In most NCSM and coupled-cluster calculations, practitioners
limit themselves to sequences of model spaces with either even or odd
values of $N$.

\section{Extrapolations and the coupled-cluster method}\label{sec:results}

In this section, we apply IR extrapolations to the results of the
coupled-cluster method with singles and doubles (CCSD)
calculations~\cite{kuemmel1978,bishop1991,bartlett2007,hagen2013c}.
Our interest is twofold. First, we want to study in more detail
whether $\Leff$ is also the relevant length scale when
\emph{approximate} many-body solutions (such as CCSD) are employed. We
note that most {\it ab initio} methods that presently compute nuclei
beyond the $p$ shell employ approximate solutions of the many-body
problem~\cite{hagen2008,barbieri2009,tsukiyama2011,hergert2013b,soma2013}.
Second, we want to probe the extrapolation formula~(\ref{master2})
over a large energy range and see how large the model space needs to
be for a reliable and accurate extrapolation.

We compute the ground-state energies of the nuclei $^{16,22,24}$O using
the next-to-next-to-leading order chiral nucleon-nucleon interaction 
of Ref.~\cite{ekstrom2014} with a cutoff
$\Lambda_\chi=500$~MeV. This interaction has been optimized to
scattering data of the nucleon-nucleon system and deuteron bound-state
properties. It is similar in quality to the chiral nucleon-nucleon
interaction NNLO$_{\rm opt}$~\cite{ekstrom2013}, which was optimized
with respect to phase shifts.

\begin{figure}[htb]
\includegraphics{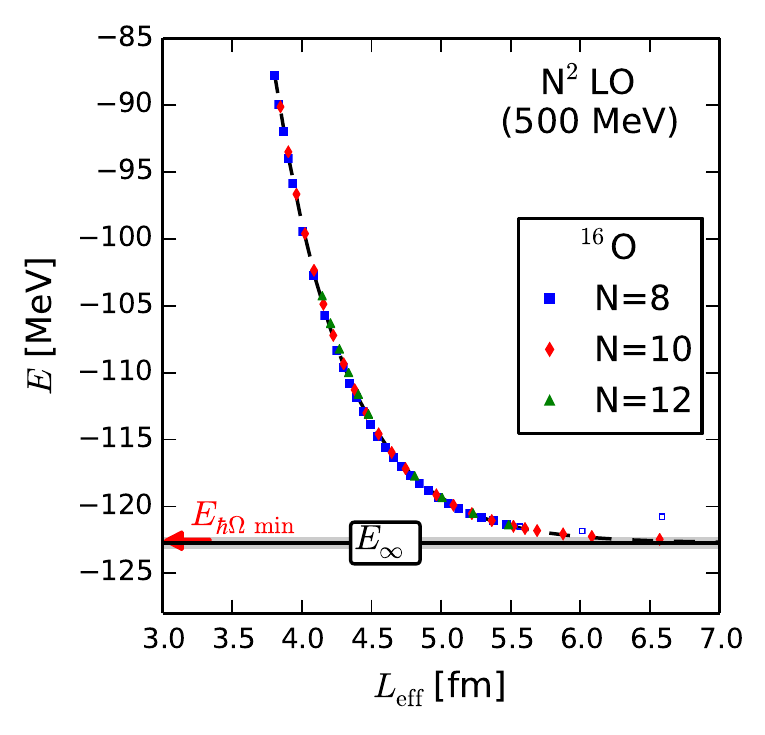}
\includegraphics{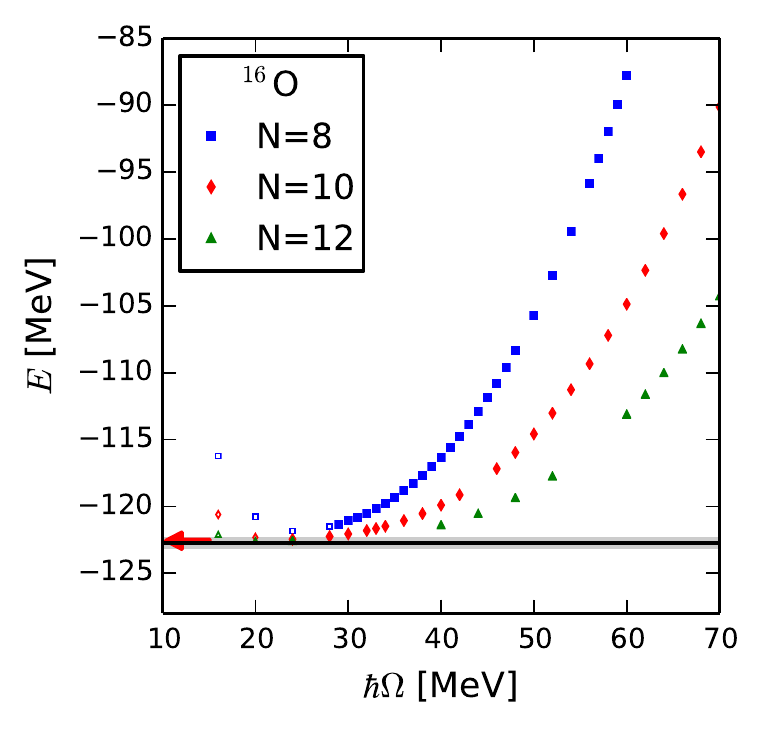}
\caption{Ground-state energies in CCSD approximation for $^{16}$O as a
  function of $\Leff$ (left panel) and $\hbar\Omega$ (right panel) for
  harmonic-oscillator spaces with $N$ as indicated. Dashed line:
  exponential fit of Eq.~(\ref{master}) to data with $N_{\rm max}=12$.
  Full line with band: asymptote $E_\infty$ from fit with $95\%$
  confidence interval.  Hollow markers: points excluded from fit. The
  arrow marks the minimum energy $E_{\hbar\Omega\,{\rm min}}$ that is
  obtained when varying the oscillator spacing $\hbar\Omega$ for
 $N_{\rm max}=12$.}
  \label{fig2} 
\end{figure}

Figure~\ref{fig2} shows the CCSD ground state energies for $^{16}$O as
a function of $\Leff$ (left figure) and $\hbar\Omega$ (right figure)
for model spaces with $N=8,10,12$.  Our model spaces have oscillator
frequencies in the interval $\hbar\Omega/({\rm MeV}) \in [16,70]$. At
fixed $N$, the energy computed at the highest oscillator frequency
corresponds to the smallest value of $\Leff$ and exhibits the smallest
UV error. In the left panel of Fig.~\ref{fig2} the solid data points
forming the exponential envelope have negligible UV corrections and
can be used in the IR extrapolation. As seen in the right panel, these
frequencies are much larger than the naive estimate $\hbar\Omega_{\rm
  min}\approx \hbar^2 \Lambda_\chi/(m R)$ that minimizes the energy
for a nucleus of radius $R$ and interaction with cutoff
$\Lambda_\chi$~\cite{hagen2010b}. The exponential
extrapolation~(\ref{master2}) to all solid data points is shown as a
dashed line, and the asymptote $E_\infty$ as a full line.

The extrapolation results are from a least-squares penalty-function
fit with equal weighting~\cite{dobaczewski2014} of the parameters
$E_\infty$, $k_\infty$, and $A_\infty$.  We note that the exponential
extrapolation practically works over the entire range of about 60~MeV.
While the energy correction can be a substantial fraction of the total
binding energy, we have $\exp{(-2k_\infty \Leff)}\ll 1$ over the
considered range of $\Leff$.  For fixed $N$, the (very small) UV
corrections are expected to increase with increasing $\Leff$, see
Fig.~20 of Ref.~\cite{more2013} as an example. We note that all solid
data points employed in the IR extrapolation in Fig.~\ref{fig2} are
from model spaces with $\Lambda_{\rm UV}> 750\mev$.  The UV cutoff
parameter of the nucleon-nucleon interaction used here is
$\Lambda_\chi\approx 500\mev$, but this is not a sharp cutoff, which
is why $\Lambda_{\rm UV}$ must be sufficiently higher than
$\Lambda_\chi$ so that effects from the omitted UV tail are small.

Let us discuss errors and error estimates. We applied a theory,
derived for the deuteron and $k_\infty \Leff\gg 1$ to many-body
systems and for $k_\infty \Leff>4$ or so. We view the extrapolation
formula~(\ref{master2}) as a leading order and systematically
improvable result applied to a complex nucleus over a wide range of
$k_\infty \Leff$, neglecting higher-order corrections.  We also deal
with systematic errors from the CCSD approximation
 (recall that $\Leff$ was worked out for an exact solution
of the operator $\sum_{i=1}^A p_i^2$). We believe that UV errors are
negligible.  Thus, we have systematic errors from neglected
corrections in the IR and from the approximate many-body method. Both
systematic errors are hard to quantify. Some aspects of these
systematic errors behave as statistical errors. For instance, we deal
with a relatively small scatter of our data points around the
exponential extrapolation. These (relatively small) errors are easy to
quantify and usually returned by fitting routines in the form of
standard asymptotic errors computed from the covariance matrix. At
this moment, these are the only errors we quantify and present in
tables below. However, we emphasize that these errors are presumably
much smaller than the systematic errors.

We fit Eq.~(\ref{master2}) to $^{16}$O, including increasingly larger
sets of data points from model spaces with $N\le N_{\rm max}$.
Table~\ref{fitO16} shows the results.  The error estimates are from
the covariance matrix. We repeat that any systematic errors from
sub-leading IR corrections or due to the CCSD approximation are not
included. We note that 
the $E_\infty$ result for $\Nmax=8$ is within the error estimates for $\Nmax=10,12$. 
Table~\ref{fitO16} also shows the minimum energy $E_{\hbar\Omega\,{\rm
    min}}$ that is obtained at fixed $\Nmax$ when varying the
oscillator spacing $\hbar\Omega$.  For the smaller model space with
$\Nmax=8$, the extrapolated energy $E_\infty$ is much closer to
the ``true'' result (from extrapolations in larger model spaces) than
the minimum energy $E_{\hbar\Omega\,{\rm min}}$, and this is the
practical value of IR extrapolations. Of course, the challenge remains
to give a meaningful error estimate for all model spaces.

\begin{table}[hbtp]
\begin{center}
  \begin{tabular}{l D{,}{\,\pm\,}{-1} D{,}{\,\pm\,}{-1} D{,}{\,\pm\,}{-1} }
    \hline\hline 
    \multicolumn{1}{c}{$\Nmax$} & \multicolumn{1}{c}{8} & \multicolumn{1}{c}{10} & \multicolumn{1}{c}{12} \\ \hline
$E_{\hbar\Omega\,{\rm min}}$ [MeV]  & -121.83     &  -122.46      &  -122.56      \\
$E_{\infty}$ [MeV]                  & -122.62,0.06 &  -122.68,0.35 &  -122.73,0.35 \\
$k_{\infty}$ [fm$^{-1}$]            &    1.00,0.00 &     0.99,0.01 &     0.98,0.01 \\
$A_{\infty}$ [$10^4$MeV]            &    6.95,0.20 &     6.48,0.49 &     5.96,0.63 \\
    \hline\hline
  \end{tabular}
\end{center}
\caption{Extrapolation parameters (and statistical error estimates)
  and energy minima for $^{16}\textrm{O}$ as a function of basis
  truncation $\Nmax$. The neglected systematic errors are expected to dominate the error budget.}
  \label{fitO16}
\end{table}

\begin{figure}[htb]
\begin{center}                                
\includegraphics{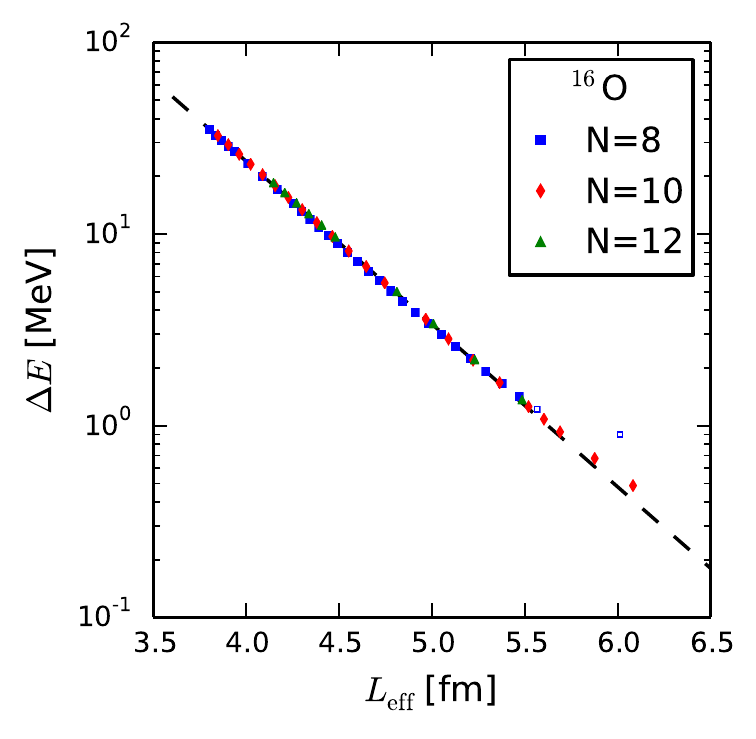}
\includegraphics{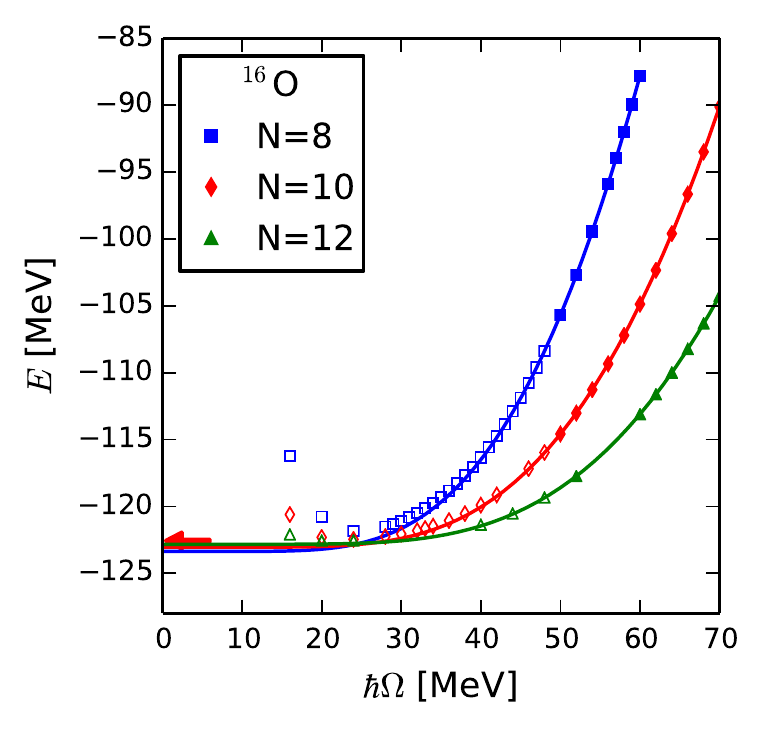}
\end{center}
\caption{Left: energy difference $\Delta E$ for the ground-state
  energy (in CCSD approximation) of $^{16}$O as a function of $\Leff$
  in a log plot. The symbols are as in Fig.~\ref{fig2}, and the dashed
  line is the exponential fit for $\Nmax = 12$ from
  Table~\ref{fitO16}.  Right: ground-state energies as a function of
  $\hw$ as in the right panel of Fig.~\ref{fig2}.  The solid lines are
  fits of Eq.~\eqref{master2} for fixed $N$ to only the solid points.
  The arrow marks the minimum energy $E_{\hbar\Omega\,{\rm min}}$ that
  is obtained when varying the oscillator spacing $\hbar\Omega$ for
  $N_{\rm max}=12$.}
  \label{fig:O16log} 
\end{figure}

The left panel of Fig.~\ref{fig:O16log} shows a log plot of the
difference $\Delta E=E-E_\infty$ for $^{16}$O. The dashed line is the
$\Nmax = 12$ exponential fit from Table~\ref{fitO16}.  Deviations at
the largest values of $\Leff$ reflect UV corrections (open symbols are not
used in the fit) and other systematic errors discussed above.
However, the consistency of the fit to Eq.~(\ref{master2}) over a
large range of $\Delta E$ is striking.  This suggests that there is
enough information for reliable extrapolations even when using only
calculations far from the energy minimum.  This is validated for
$^{16}$O in the right panel of Fig.~\ref{fig:O16log}, where
extrapolations are shown for fixed $N=8$, $10$, and $12$ using only
points in each case with $\hw \ge 50\,$MeV.  This plot can also be
interpreted as showing that controlled and consistent (if not highly
precise) extrapolations to $\hw = 0$ (i.e.\ removing the IR cutoff for
a fixed $N$) can be achieved if UV corrections are suppressed.

\begin{figure}[tb]
\includegraphics{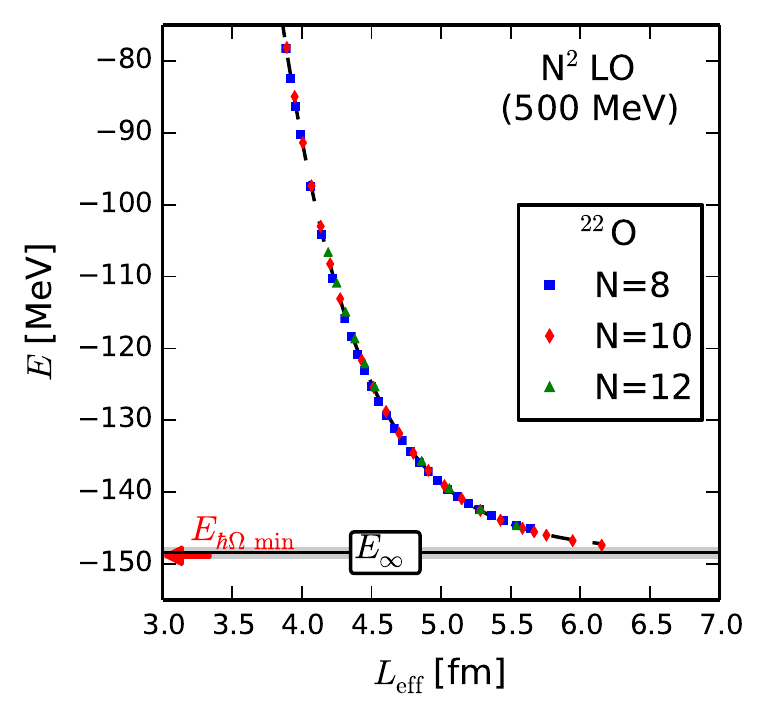}
\includegraphics{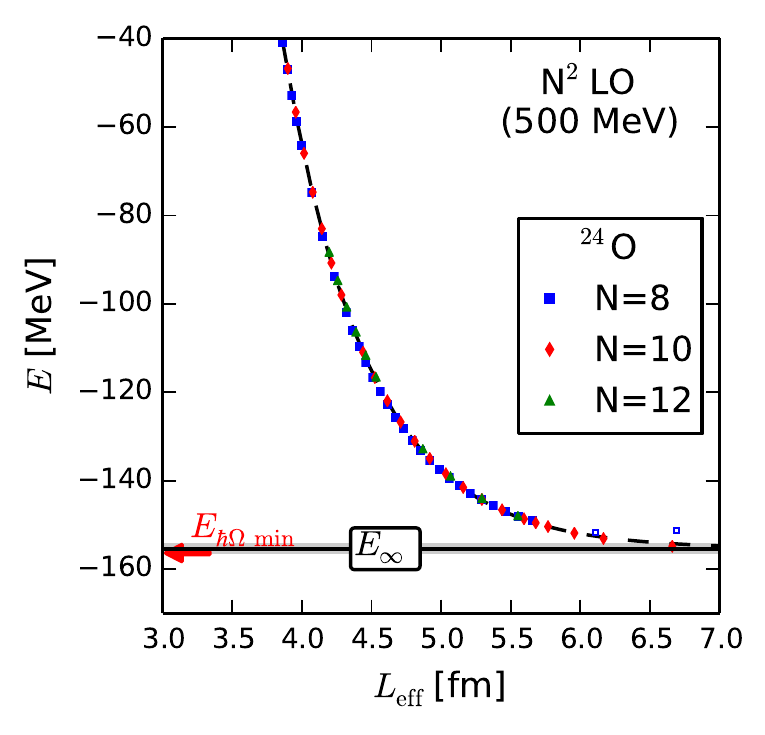}
\caption{(Color online) Ground-state energies (CCSD approximation) for
  $^{22}$O (left) and $^{24}$O (right) as a function of $L_2$ for
  harmonic oscillator spaces with $N$ as indicated. Dashed line:
  Exponential fit to Eq.~(\ref{master2}). Full line with band:
  Asymptote $E_\infty$ from errors from covariance matrix.  Hollow
  markers: Points excluded from fit. The arrow marks the minimum
  energy $E_{\hbar\Omega\,{\rm min}}$ that is obtained when varying
  the oscillator spacing $\hbar\Omega$ for $N_{\rm max}=12$.}
  \label{fig3} 
\end{figure}

\begin{figure}[tb]
\begin{center}                                
\includegraphics{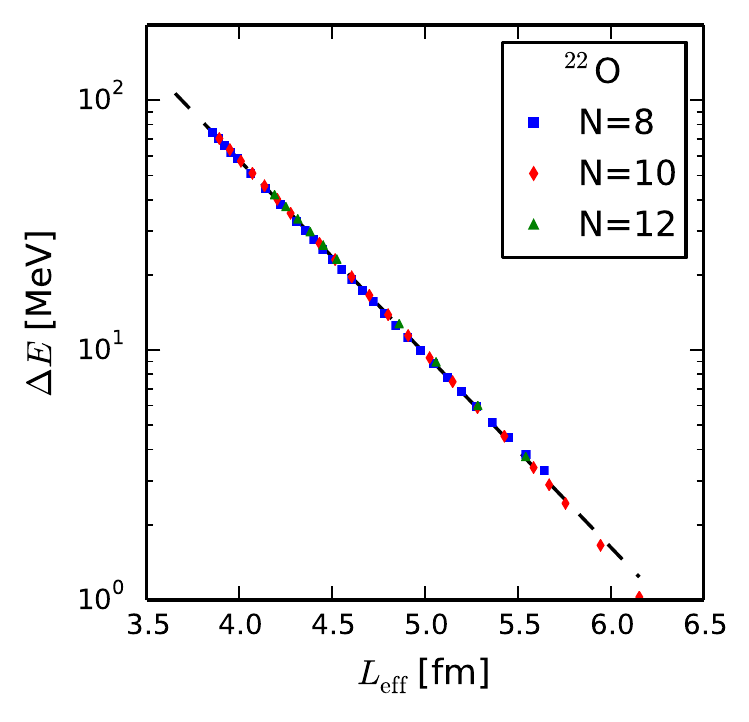}
\includegraphics{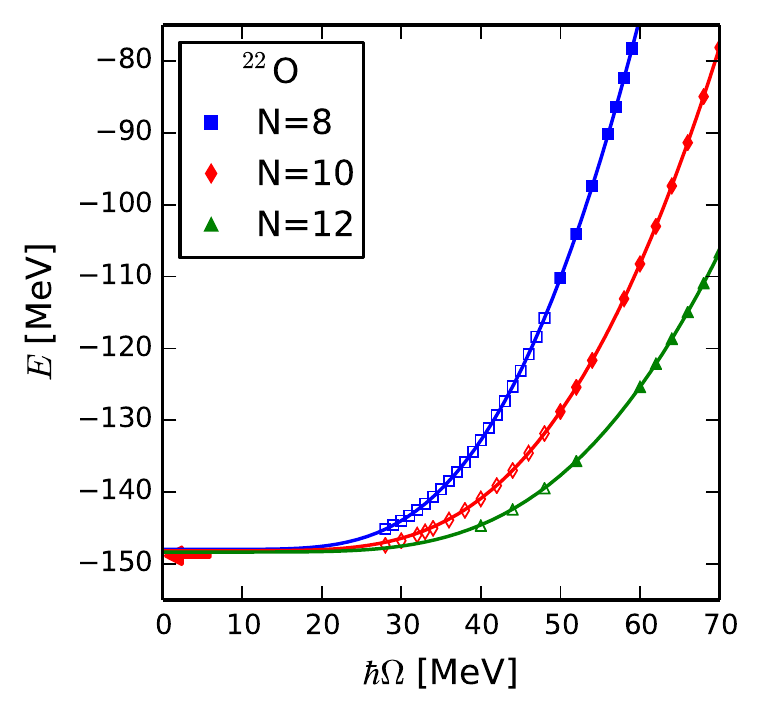}
\end{center}
\caption{Left: energy difference $\Delta E$ for the ground-state
  energy (in CCSD approximation) of $^{16}$O as a function of $\Leff$
  in a log plot. The symbols are as in Fig.~\ref{fig2}, and the dashed
  line is the exponential fit for $\Nmax = 12$ from
  Table~\ref{fitO2224}.  Right: ground-state energies as a function of
  $\hw$ as in the right panel of Fig.~\ref{fig2}.  The solid lines are
  fits of Eq.~\eqref{master2} for fixed $N$ to only the solid points.
  The arrow marks the minimum energy $E_{\hbar\Omega\,{\rm min}}$ that
  is obtained when varying the oscillator spacing $\hbar\Omega$ for
  $N_{\rm max}=12$.}
  \label{fig:O22log} 
\end{figure}

\begin{figure}[htb]
\begin{center}                                
\includegraphics{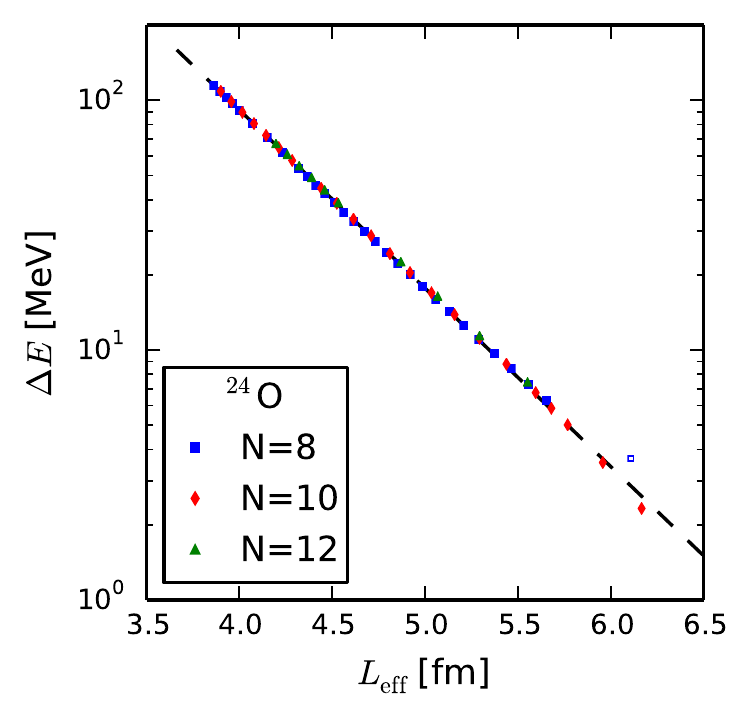}
\includegraphics{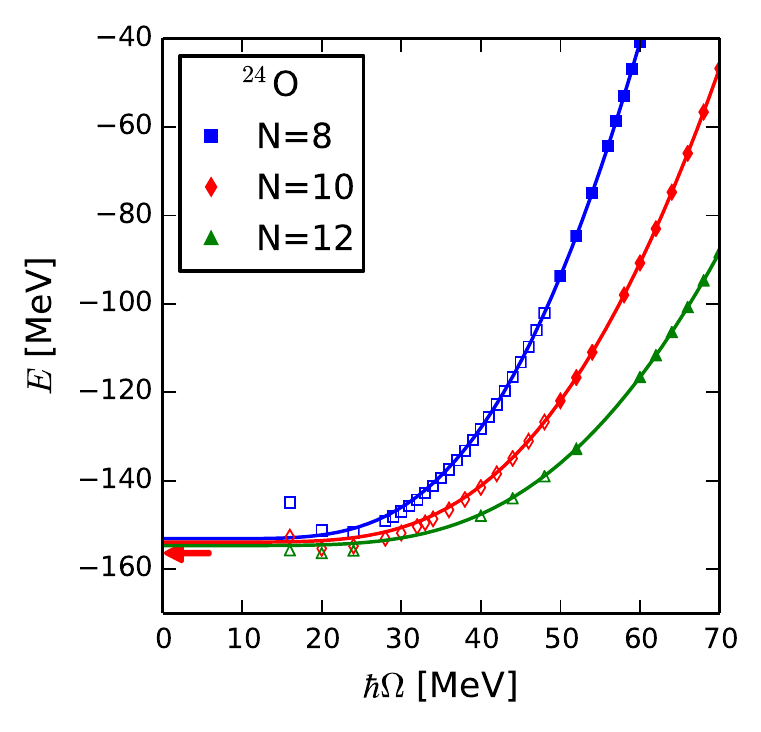}
\end{center}
\caption{Left: energy difference $\Delta E$ for the ground-state
  energy (in CCSD approximation) of $^{24}$O as a function of $\Leff$
  in a log plot. The symbols are as in Fig.~\ref{fig2}, and the dashed
  line is the exponential fit for $\Nmax = 12$ from
  Table~\ref{fitO2224}.  Right: ground-state energies as a function of
  $\hw$ as in the right panel of Fig.~\ref{fig2}.  The solid lines are
  fits of Eq.~\eqref{master2} for fixed $N$ to only the solid points.
  The arrow marks the minimum energy $E_{\hbar\Omega\,{\rm min}}$ that
  is obtained when varying the oscillator spacing $\hbar\Omega$ for
  $N_{\rm max}=12$.}
  \label{fig:O24log} 
\end{figure}

Now we turn to the neutron-rich isotopes $^{22,24}$O. Figure~\ref{fig3}
shows the CCSD ground-state energies for $^{22}$O (left) and $^{24}$O
(right) as a function of $\Leff$. The exponential
extrapolation~(\ref{master2}), and the extrapolated ground-state
energies $E_\infty = -148.41 \pm 0.60\,\mbox{MeV}$ (for $^{22}$O) and
$E_\infty=-155.38 \pm 0.83\,\mbox{MeV}$ (for $^{24}$O) are also shown.
These results employ model spaces with $N=8,10,12$.
Table~\ref{fitO2224} shows the extrapolated energies (and error
estimates from the $\chi^2$ fit) when only data points with $N\le
N_{\rm max}$ are employed in the fit. 
We note that the fits work well over an energy range of tens of
\,MeVs (see the left panels of Figs.~\ref{fig:O22log} and \ref{fig:O24log}). 
Again, the value of the IR
extrapolation lies in the finding that $\Nmax=8$ extrapolations
yield results that are close to the ``true'' ground-state energies
(see the right panels of Figs.~\ref{fig:O22log} and \ref{fig:O24log}).

\begin{table}[hbtp]
\begin{center}
  \begin{tabular}{c l D{,}{\,\pm\,}{-1} D{,}{\,\pm\,}{-1} D{,}{\,\pm\,}{-1}}
    \hline\hline 
    & \multicolumn{1}{c}{$N_{\textrm{max}}$} & \multicolumn{1}{c}{8} & \multicolumn{1}{c}{10} & \multicolumn{1}{c}{12}\\ \hline
    \multirow{3}{*}{$^{22}\textrm{O}$} & 
      $E_{\infty}$ [MeV]        & -147.93,0.01 & -148.27,0.47 & -148.41,0.60 \\
    & $k_{\infty}$ [fm$^{-1}$]  &    0.91,0.00 &    0.90,0.01 &    0.89,0.00 \\
    & $A_{\infty}$ [$10^4$MeV]  &    8.56,0.02 &    7.76,0.52 &    7.19,0.33 \\
\hline
    \multirow{3}{*}{$^{24}\textrm{O}$} & 
      $E_{\infty}$ [MeV]        & -154.42,0.10 & -155.08,0.63 & -155.38,0.83 \\ 
    & $k_{\infty}$ [fm$^{-1}$]  &    0.83,0.00 &    0.83,0.01 &    0.82,0.00 \\
    & $A_{\infty}$ [$10^4$MeV]  &    7.89,0.11 &    7.06,0.52 &    6.53,0.31 \\
    \hline\hline
  \end{tabular}
\end{center}
\caption{Extrapolated energies $E_\infty$ for $^{22,24}\textrm{O}$ as
  a function basis truncation $N_{\textrm{max}}$.  The energy minima
  for $^{22,24}\textrm{O}$ in a $N_{\rm max}=12$ model-space are found at the oscillator
  frequency $\hbar\Omega = 20\mev$ and are $E_{\hbar\Omega~{\rm min}}
  = -148.85\mev$ and $E_{\hbar\Omega~{\rm min}} = -156.35\mev$, respectively.}
  \label{fitO2224}
\end{table}

Our study shows that IR extrapolation can be a practical tool for
approximate solutions of the nuclear many-body problem. The
extrapolation is accurate and reliable over a large energy range of
tens of MeV. For the employed chiral interaction at NNLO, the examples
of $^{16,22,24}$O suggest that ground-state energies of $p$-shell and
$sd$-shell nuclei can be extrapolated from model spaces with $N=8$.

We also tried to include $N=6$ data points in the extrapolation, but
the data points did not fall onto the same line as the $N=8,10,12$
points. We speculate that this is due to peculiarities of the CCSD
approximation, or due to a less complete decoupling of the center of
mass in small model spaces~\cite{hagen2010b}. Again, this points to
the need to better understand the systematic errors involved in IR
extrapolations of results obtained with approximate many-body results.

\section{Summary}
We studied IR extrapolations for coupled-cluster computations of
oxygen isotopes with chiral nucleon-nucleon interactions at NNLO. One
of the main results is the identification of the nucleus-dependent
infrared box size $\Leff$. Our results show that IR extrapolations are
feasible in practice, but we need a better understanding of systematic
errors for extrapolated quantities; a Bayesian framework may be useful
in this regard.  Nevertheless, we demonstrated that reliable IR
extrapolations can be performed in \emph{p}-\emph{s}-\emph{d} nuclei
spanning over tens of MeVs.  For the IR extrapolations to work in
practice one needs to minimize UV corrections and work at frequencies
away from the usual range about the oscillator frequency that
minimizes the energy at fixed number $N$ of oscillator quanta.

\ack This material is based upon work supported in part by the
National Science Foundation under Grant No.~PHY--1306250 (Ohio State
University), by the U.S. Department of Energy, Office of Science,
Office of Nuclear Physics, under Award Numbers DE-FG02-96ER40963
(University of Tennessee), DE-SC0008499/DE-SC0008533 (SciDAC-3 NUCLEI
Collaboration), the Field Work Proposal ERKBP57 at Oak Ridge National
Laboratory, and under contract number DEAC05-00OR22725 (Oak Ridge
National Laboratory).

\section*{References}

\bibliographystyle{iopart-num}

\bibliography{refs}

\end{document}